\documentclass[useAMS,usenatbib]{mn2e}
\usepackage{natbib}

\usepackage{amsmath,amsfonts}
\usepackage{epsfig}

\def\reff@jnl#1{{\rm#1\/}}

\def\aj{\reff@jnl{AJ}}                  
\def\araa{\reff@jnl{ARA\&A}}            
\def\apj{\reff@jnl{ApJ}}                        
\def\apjl{\reff@jnl{ApJ}}               
\def\apjs{\reff@jnl{ApJS}}              
\def\ao{\reff@jnl{Appl.Optics}}         
\def\apss{\reff@jnl{Ap\&SS}}            
\def\aap{\reff@jnl{A\&A}}               
\def\aapr{\reff@jnl{A\&A~Rev.}}         
\def\aaps{\reff@jnl{A\&AS}}             
\def\azh{\reff@jnl{AZh}}                        
\def\baas{\reff@jnl{BAAS}}              
\def\jrasc{\reff@jnl{JRASC}}            
\def\memras{\reff@jnl{MmRAS}}           
\def\mnras{\reff@jnl{MNRAS}}            
\def\pra{\reff@jnl{Phys.Rev.A}}         
\def\prb{\reff@jnl{Phys.Rev.B}}         
\def\prc{\reff@jnl{Phys.Rev.C}}         
\def\prd{\reff@jnl{Phys.Rev.D}}         
\def\prl{\reff@jnl{Phys.Rev.Lett}}      
\def\pasp{\reff@jnl{PASP}}              
\def\pasj{\reff@jnl{PASJ}}              
\def\qjras{\reff@jnl{QJRAS}}            
\def\skytel{\reff@jnl{S\&T}}            
\def\solphys{\reff@jnl{Solar~Phys.}}    
\def\sovast{\reff@jnl{Soviet~Ast.}}     
\def\ssr{\reff@jnl{Space~Sci.Rev.}}     
\def\zap{\reff@jnl{ZAp}}                        
\def\nat{\reff@jnl{Nature}}             

\title[Anthracene cations in the ISM]{A search for interstellar anthracene toward the Perseus anomalous microwave emission region }

\author[S. Iglesias-Groth et al.] {S.
Iglesias-Groth,$^{1,2}\thanks{E-mail: sigroth@iac.es}$ A.
Manchado,$^{1,2,3}$ R. Rebolo,$^{1,2,3}$ \newauthor J. I. Gonz\'alez
Hern\'andez,$^{1,2,4}$ D. A. Garc\'\i a-Hern\'andez$^{1,2}$ and
D.L. Lambert$^5$\\  
$^1$ Instituto de Astrofis\'{\i}ca de Canarias, 38200 La Laguna, Tenerife, Canary Islands, Spain \\
$^2$ Universidad de La Laguna, E-38205 La Laguna, Tenerife, Spain \\
$^3$ Consejo Superior de Investigaciones Cient\'{\i}ficas, Spain \\
$^4$ Dpto. de Astrof{\'\i}sica y Ciencias de la Atm\'osfera, Facultad de  Ciencias F{\'\i}sicas, Universidad Complutense de Madrid, E-28040  
Madrid, Spain\\
$^5$ The W.J. McDonald Observatory, University of Texas, Austin, TX 78712-1083, USA
}  

\date{Accepted Received In original form}
\pagerange{\pageref{firstpage}--\pageref{lastpage}}
\pubyear{2010}

\begin{document}

\label{firstpage}
\maketitle

\begin{abstract}
We report the discovery of a new broad interstellar (or circumstellar)  band at 7088.8$\pm$2.0 \AA~ coincident to within the measurement uncertainties with the strongest band of the
anthracene cation (C$_{14}$H$_{10}$$^+$) as  measured  in gas-phase  laboratory spectroscopy at  low temperatures (Sukhorukov et al.2004). The band is detected in the
line of sight of star Cernis 52, a likely member of the very young star cluster IC 348,  and is probably  associated with cold absorbing material in  a intervening
molecular cloud of the  Perseus star forming region where various experiments have recently detected anomalous microwave emission. From the measured intensity and available oscillator strength we find 
a column density of N$_{an^+}$= 1.1($\pm$0.4) x 10$^{13}$ cm$^{-2}$ implying that  $\sim$ 0.008 \%   of the carbon
in the cloud  could be in the form of C$_{14}$H$_{10}$$^+$.  A similar abundance has been recently claimed for the naphthalene cation  (Iglesias-Groth et al. 2008) 
in this cloud. This is the first  location  outside the Solar System where specific PAHs are identified.  We report observations of interstellar lines of  CH
and CH$^+$  that support a rather high   column density  for these species  and
for molecular hydrogen. The strength ratio of the two prominent diffuse
interstellar bands at 5780 and 5797 \AA~ suggests the presence of a ``zeta'' type  cloud in the line of sight (consistent with steep far-UV extinction and high molecular content).  The presence of  PAH cations and other related  hydrogenated carbon molecules which
are likely to occur in this type of clouds reinforce the suggestion that 
electric dipole radiation from fast spinning PAHs is responsible of  the anomalous microwave emission detected toward Perseus.  

\end{abstract}


\section{Introduction}

Several regions in the Perseus molecular complex display a  strong
continuum microwave emission in the frequency range 10-60 GHz   that
is correlated with dust thermal  emission (Watson et al. 2005, Tibbs
et al. 2010). 
The spectral dependence of this emission  cannot be explained by 
synchrotron, free-free or thermal dust radiation  processes and
represents one of the best known examples of the so-called anomalous
microwave emission (Kogut et al. 1996, Leitch et al. 1997, de Oliveira
et al. 1999, 2002). This kind of microwave emission is also  detected
in other molecular clouds (Casassus et al. 2006)  and there is
increasing statistical evidence supporting its  existence at high
Galactic latitudes (Hildebrandt et al. 2007). Draine and 
Lazarian (1998) suggested a possible explanation for this  emission   based on
electric dipole radiation from rapidly spinning  carbon based molecules (see
also Iglesias-Groth 2005, 2006).  Polycyclic aromatic hydrocarbons (PAHs)
proposed as the sources of interstellar mid-infrared emission (Puget \& L\'eger
1989, Allamandola, Tielens \& Barker 1989) and diffuse interstellar
bands (Puget \& L\'eger 1989, Allamandola, Tielens \& Barker 1989) are
 also potential  carriers for the anomalous  microwave emission.  
It is  likely that anomalous microwave emission surveys reveal
interstellar  regions of enhanced PAH abundance and therefore further
studies of these regions may  facilitate the identification of
individual PAHs.  
Recent progress in the laboratory measurements of optical and near
infrared bands of the most simple PAHs and their cations under
conditions resembling  the interstellar medium  is a key step  to this
aim (e.g. Salama 2008).  

Here, we report results on a  search  for interstellar bands  in the
line of sight of the maximum anomalous microwave emission detected
toward the  Perseus clouds (Tibbs et al. 2010). We obtained high
resolution 
optical spectroscopy of  Cernis 52 (BD+31$^o$ 640), a reddened star
(E(B-V)=0.9, Cernis 1993) located in this line of sight at a distance
of 240 pc where most of the dust extinction is in the Perseus OB2 
dark cloud complex. This is the brightest hot star that we could 
identify as a suitable background object to study the interstellar
absorptions in the intervening cloud.  Very recently, evidence was
reported for absorption bands of  the naphthalene cation
C$_{10}$H$_8$$^+$  in the line of sight of this star (Iglesias-Groth
et al. 2008). A detailed chemical composition study of this star has
been  published by Gonz\'alez Hern\'andez et al. (2009).  We report
now  evidence in this line of sight for the next PAH in terms of
complexity, 
the anthracene  molecule which is composed of three aromatic rings.
Laboratory spectroscopy of the anthracene cation
(C$_{14}$H$_{10}$$^+$)  obtained using supersonic jet-discharge cavity
ring-down 
spectroscopy (Sukhorukov et al. 2004) to resemble the conditions of
the interstellar medium has  provided an  accurate measurement of the 
 strongest vibronic band for the D$_2$$\leftarrow$D$_0$   system 
of this molecule. The laboratory band is measured at the wavelength of
7085.7 $\pm$ 1.3 \AA~ (in standard air) with a full width at half
maximum of $\sim$ 47 \AA~. We obtained spectra  of  Cernis 52 and
found  a new broad interstellar feature   with wavelength and width 
consistent with the   measurements for the strongest optical
transition of the anthracene cation.  We also present new measurements
of  interstellar absorption lines of  CH and CH$^+$ and of the most
prominent diffuse interstellar bands  providing  insight on the
physical and chemical processes in the intervening medium. 

\begin{figure*}
\includegraphics[angle=0,width=8cm,height=8cm]{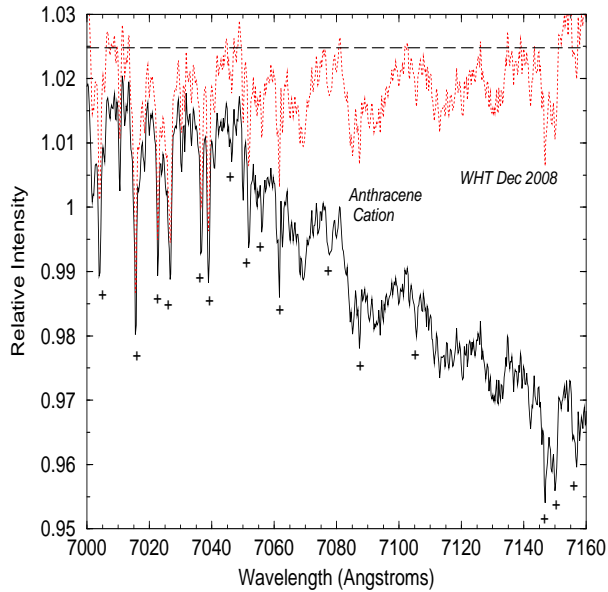}
\caption{Spectrum of Cernis 52 obtained at WHT with ISIS (grating
R1200R)  on December 2008 before (solid black line)  and after  (red
dotted line) normalization of the continuum.   The strongest telluric
lines are marked. The dashed line indicates the location of the
continuum.} 
\label{fig:f1}
\end{figure*}

\section{Observations}

We report spectroscopy of  Cernis 52  obtained with: a)  the ISIS
spectrograph (grating R1200B and R1200R  for the blue and red arm,
respectively) at the 4.2 m William Herschel Telescope (WHT), at the
Roque de los Muchachos Observatory (La Palma, Spain) on  December 2007
and December 2008; b)  the High Resolution Spectrograph (HRS) of the
9.2 m  Hobby-Eberly Telescope at McDonald Observatory (Texas, USA) on 
several nights during 2008.  
  
Comparison  stars were observed with the same instruments on the same
nights for correction of telluric lines and control of the
instrumental response.  All the spectra were reduced using IRAF,
wavelength calibrated with less  than 20 m\AA~ rms error,  and
individual spectra were co-added  after   correction for  telluric
lines.  The adopted  slit width for ISIS  (1.00 arcsec) led to a
spectral resolution of 0.8  and 0.6 \AA~ for the blue and red arm of
ISIS, respectively.  The fiber used to feed  HRS  led to a spectral
resolution of  025 \AA in the spectral regions of interest.   The
zero-point of the adopted wavelength scale   was set by assigning the
laboratory wavelength to the KI 7698.974 \AA~ interstellar absorption
line. We plot in Fig. 1 one of the spectra of Cernis 52 obtained at
the WHT. This spectrum was extracted using the IRAF routine ``apsum''
after bias and flatfield correction of the original spectral image. We
show the spectrum with no normalization  applied, only division by a
constant and we also show the result of normalization using a low
order polynomial fit.  We note the smooth  behaviour of the continuum
and the presence of several broad absorption features and narrow
telluric lines. The continuum correction does not affect these
features. The WHT spectra  help to define 
the continuum regions used for normalization of the echelle spectra 
obtained at HET. 

The  Cernis 52 spectra from both telescopes were  corrected for
telluric line contamination and possible instrumental effects dividing
each individual spectrum  (15 were recorded at HET)   by the almost
featureless spectrum of  a much  brighter, hot and fast rotating star
in a nearby line of sight observed with the same instrument
configuration. After this correction, individual spectra  were
normalized using a  polynomial fit to the continuum regions and then 
combined to improve S/N. The final  WHT and HET spectra  are plotted
at the top of  Fig. 2 (red and blue solid lines, respectively).  We
remark  the good agreement between the WHT and HET spectra (both with 
S/N $\geq$ 300 per pixel).  These two  spectra contain stellar
photospheric features   as well as   bands associated with 
circumstellar or diffuse interstellar material in the line of sight
(Iglesias-Groth et al. 2008, Gonz\'alez Hern\'andez et al. 2009 ).
Cernis 52 is embedded in a cloud responsible for significant visible
extinction (A$_V$$\sim$3) and the presence of  absorption  bands
caused by the intervening interstellar material were therefore 
expected.

\begin{figure*}
\includegraphics[angle=0,width=11cm,height=11cm]{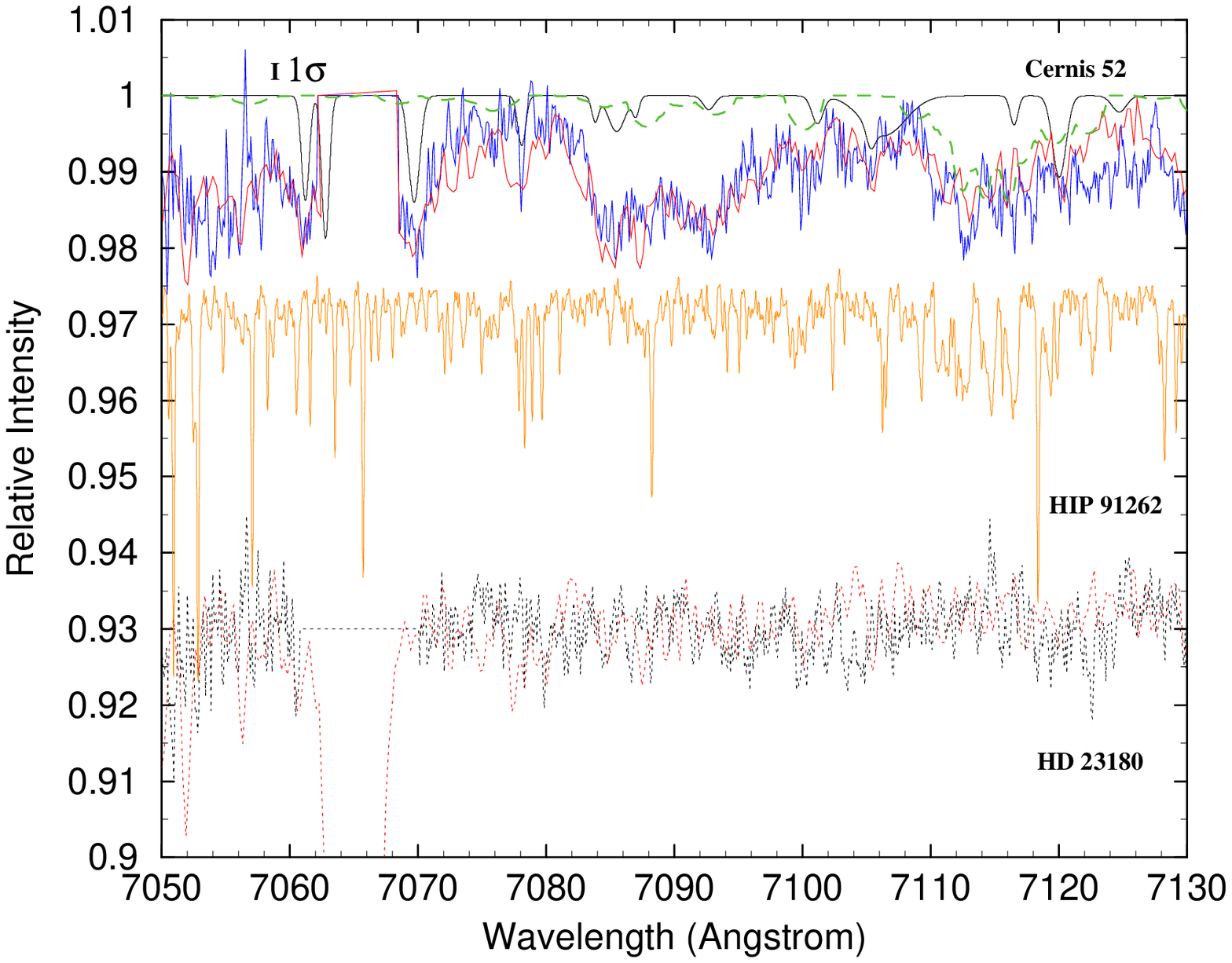}
\caption{Top. Final spectra of  Cernis 52 (WHT: red line; HET: blue
line) showing two independent detections of the new 7088.8 \AA~ band
ascribed to the anthracene cation. 1$\sigma$  error bars are
indicated. Both spectra have been corrected for telluric lines using 
rapidly  rotating hot stars observed with the same instrumental
configuration.  Telluric absorptions were divided out except in the
range 7060-7070 \AA~  where a He line present (see spectrum at the
bottom) prevents this correction. The uncorrected spectral region has 
been set at unity. A photospheric spectrum (dashed green line)
computed  for the stellar parameters of Cernis 52 (see text for
details)  and  a DIB synthetic spectrum (solid thin black line)  based
on data from Hobbs et al. (2008, 2009) are overplotted. Middle.
Spectrum of the comparison AO-type star HIP 91262 (thin red line) from
Allende-Prieto et al. (2004). The narrow absorptions indicate the
location of telluric lines which were largely corrected for Cernis 52.
 Bottom. Spectra  of the reference hot star HD 23180 (WHT: red line;
 HET: dotted black line) observed with the same instrumentation than
 Cernis 52. The HET spectrum was divided by  the  same fast rotator
 used  for Cernis 52.  No  division was applied to the  WHT data in
 order to show the He 7065 \AA~ absorption and the strength of  
 telluric lines.}  
\label{fig:f2}
\end{figure*}

\section{Results and Discussion}

In both  spectra of Cernis 52 plotted in  Fig. 2 we note the presence of  a broad absorption in the range 7075-7105 \AA. This is the range  where absorption from the  most intense band of the anthracene cation is  expected.   As we can see there is no similar absorption feature   in the  spectrum of the nearby hot star HD 23180 (Cernis 67, also known as o Per)   plotted at the bottom. This star is located in  Perseus close but outside the region of 
anomalous microwave emission.   It  was  observed with the same instruments and with similar  signal to noise ratio to Cernis 52 at both
telescopes. The HD 23180 and the Cernis 52 spectra were   divided by the same spectra of fast rotating stars to correct for telluric   absorptions.
The rather flat  spectra of HD 23180 indicate that the broad  absorption feature in the range 7075-7105 \AA~ found in Cernis 52 is not an 
artefact introduced by the instrument or the reduction  procedure.  In Fig. 2 we can see the absence of broad  absorptions in the spectrum of the A0V star HIP 91262 taken from the database by Allende-Prieto et al. (2004). The very narrow absorptions  are telluric lines. The most significant photospheric absorption in this star  is located in the range 7110-7120 \AA~ where we also see absorption in the spectrum of Cernis 52. This is mostly caused by  photospheric  C I lines.

In what follows we show in more detail that the broad absorption in the range 7075-7105 \AA~ does not originate in the photosphere of Cernis 52. We
computed  a synthetic spectrum of Cernis 52 using  the atmospheric parameters and metallicity derived for this star by Gonz\'alez Hern\'andez et al. (2009). We
adopted T$_{eff}$ = 8350 K, log (g/cm s$^{-2}$)=4.2 and [Fe/H]=-0.01. The synthetic spectrum was broadened to match  the rotational velocity  of the star (v sin i
= 65 km/s). In Gonz\'alez Hern\'andez et al.  evidence is reported  for veiling of the photosphericspectrum,  they find that almost half of the total observed
flux at 5180 \AA~ is contributed by the photosphere of Cernis 52. We also find that the synthetic spectrum has to be veiled by an amount comparable to the flux
in the continuum  in order  to match the photospheric features in the 7110-7120\AA~ region. The veiled synthetic spectrum is plotted at the top of  Fig. 2
(dashed green line) where we can  see there are only very weak  photospheric absorptions in the range 7050-7110 \AA~ which  cannot explain the broad feature
detected in the spectrum.

As an additional check, we compared our synthetic spectra with  stellar spectra available in the compilation by 
Allende-Prieto et al. (2004)  and  verified that the photospheric features  of  stars with spectral type similar to Cernis 52 were properly
reproduced. To illustrate this, we plot in Fig. 3 the  synthetic spectrum computed for Cernis 52  and compare it with the
spectrum of the A3 V star HIP 57632 (corrected for telluric lines) and broadened to the rotational velocity of Cernis 52.  The reference star shows no 
significant photospheric absorptions in the 7075-7105 \AA~ range.  The
main photospheric  absorption is the C I  blend at 7110-7120 \AA~
which is reasonably well reproduced in the computed spectrum.  We
trust  our spectral synthesis can reproduce the observed features in
the spectra of stars similar to Cernis 52 and  therefore conclude 
that  photospheric lines cannot explain the broad absorption detected
in the range 7075-7105 \AA. We find other possible broad  bands in
Cernis 52   at 7050 and  7133 \AA~  which are  not present in the
reference stars and cannot  be abscribed to known  photospheric or
interstellar features. The TiO $\gamma$ system has bandheads at 7054,
7088 and 7125 \AA~ but the atmospheric  temperatures  of both
components of Cernis 52 are much higher than required for  the
formation of atmospheric TiO molecules   and therefore cannot provide
an explanation. The  potential contribution to these absorptions of 
circumstellar or protoplanetary  disk  material at  much cooler
temperatures  cannot be ruled out neither proved at present and should
be futher investigated.    

\begin{figure*}
\includegraphics[angle=0,width=8cm,height=8cm]{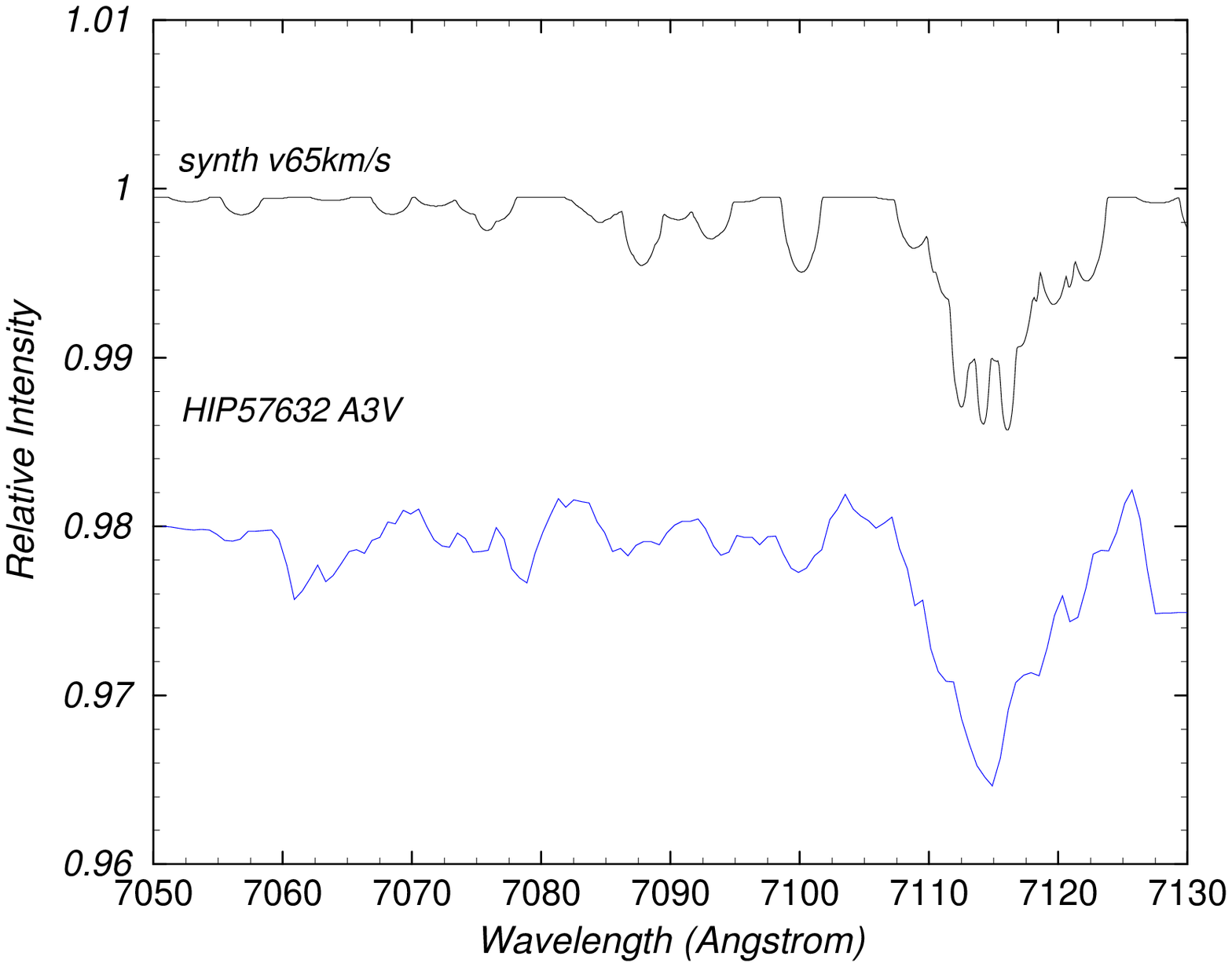}
\caption{Top: Synthetic spectrum computed for the stellar parameters
of Cernis 52. Bottom: Spectrum of the star HIP 57632 (A3 V) from
Allende-Prieto et al. (2004) box-car smoothed to match the rotational
velocity of Cernis 52. Telluric lines are corrected.} 
\label{fig:f3}
\end{figure*}

In Fig. 2 we  also overplot a synthetic spectrum of DIBs  computed
adopting wavelengths,  widths and  strengths from a  comprehensive 
compilation  in spectra  of the  stars HD 204827  and HD 183143 (Hobbs
et al. 2008, 2009). These stars display extinctions E(B-V) of 1.11 and
1.27, respectively, slightly higher than Cernis 52 and   provide a
valuable    reference for the strength of the DIBs in our target. The
assumption of a  gaussian profile was reasonable, in particular given
the relatively modest  resolution of our spectra,  for most of the
DIBs in the wavelength range of interest.  
For DIBs in common among these two works  we adopted the  equivalent widths by Hobbs et al. (2009) which are of slightly higher quality but also
correspond to a higher extinction. Very few  DIBs where present only in one of the two reference stellar spectra.  We can  identify in the
spectrum  of Cernis 52 absorption features coincident with  previously  known DIBs  located at 7069.68, 7078.07, 7083-87, 7092.67 and 7106.31 \AA.
For these DIBs we measure   equivalent widths  of 7, 5, 10, 6  and $\sim$ 28 m\AA, respectively (with  errors of order 10 \%) which seem to be
weaker in Cernis 52  than in the reference stars.   We  convolved the synthetic DIB spectrum to account for the resolving power of our observed
spectra. Then, we scaled the DIB spectrum  to provide  a reasonable  match to the DIBs in Cernis 52 that were located	outside  the region
of the anthracene. In practice we  establish the scaling factor from
the ratios of equivalent widths for several DIBs adjacent to the
anthracene region. We  find that the equivalent widths of the DIBs in
our spectra compare well in strength with those in HD 204827, while a
reduction of $\sim$30\% is required for the DIB spectrum provided by
HD 183143. The synthetic DIB  spectrum with this factor applied is
plotted  at the top of Fig. 2.  We can see that the DIB spectrum
reproduces reasonably well several absorptions in our spectra of
Cernis 52, in particular the stronger DIBs  at 7061, 7063, 7070,  and
7120 \AA. It is also clear that the DIB spectrum cannot explain the 
broad feature in the 7075-7105 \AA~ range. We conclude that this
feature in the spectrum of Cernis 52 cannot be of photospheric origin
since it is at lest 10 times broader and much stronger  than
photospheric lines and cannot be explained by known DIBs which are 
10 times narrower and much  weaker in this spectral range. To our
knowledge, no such broad band	has been reported in spectra of
similar stars, so we suggest that it could be associated with
molecular material of circumstellar  or interstellar origin. 

In order to better quantify the shape and strength of a potential anthracene
cation feature, we have applied corrections  for both the photospheric bands and the DIBs
present in the spectra. First,  we removed  the photospheric lines from the
spectra of Cernis 52  by dividing by the synthetic  photospheric spectrum of
Fig. 2. Then, the resulting spectrum was divided by either the synthetic DIB
spectrum  of Fig. 2   or by the publicly available  "average" spectrum of stars
of comparable extinction to Cernis 52  (Galazutdinov et al. 2000). These
corrections were applied to the spectra of Cernis 52 obtained at both telescopes.
 The results obtained with each of the two  DIB reference spectra are
plotted in Fig. 4 where we can see that  independent from  the telescope 
(WHT or HET)  and from the DIB reference spectrum used for  correction (the
synthetic spectrum based on  data by Hobbs et al. or the Galazutdinov et al.
``average'' spectrum)   we obtain as a result   a  broad band remarkably close
in wavelength and width   to the laboratory measurements of  the strongest band
of the anthracene cation. Spectra ``b'' and ``c'' allow us to compare 
the use of a ``synthetic'' DIB spectrum versus a real one. The latter
may describe more properly the profile of the bands but may also
introduce 
photospheric contamination from the  hot reference star. As we can
see,  using one approach or the other does not make  a significant
difference for the shape and strength of the  band we are interested.
Apparent minor differences in the resulting spectra can be attributed
to uncertainties in the  scaling and differential behaviour of DIBs
with E(B-V) and to the presence of small atmospheric features in the
hot stars used to outline these DIBs. 
We  caution   that the "average spectrum" used as
divisor  can introduce features which are difficult to trace. It is unfortunate
that the blue wing of our band is affected by a He line present in hot stars
(see HD 23180 spectra in Fig. 2)  and certainly  in the "average reference
spectrum" limiting our ability to obtain suitable  corrections at a wavelength
bluer than 7070 \AA. This region is also closer to the edge  of the order in
the  HET echelle  data where the S/N decreases quickly. 

\begin{figure*}
\includegraphics[angle=0,width=11cm,height=11cm]{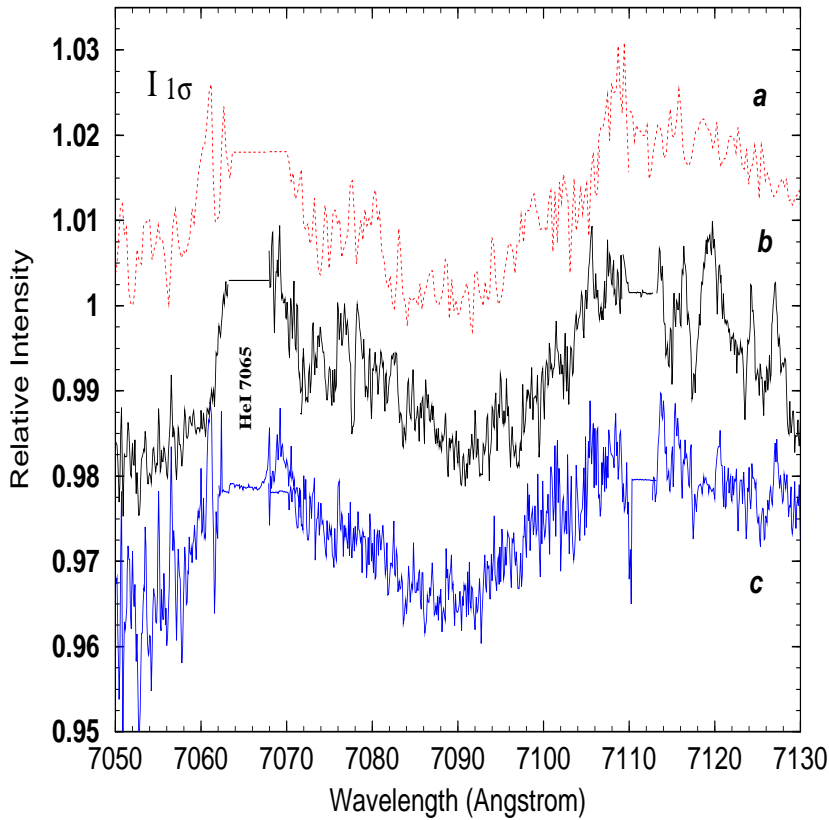}
\caption{ HET and WHT spectra of Cernis 52 corrected for the
photospheric and DIB contributions. The red solid line shows the 
results obtained for the WHT spectrum (a) corrected using the
``average''  DIB spectrum  of Galazutdinov et al. (2000). The black
solid line shows the HET data (b) corrected using the DIB spectum in
Fig.2.  The blue solid line shows the results for the HET data (c) 
corrected using the ``average'' DIB spectrum. All the spectra  have
been normalized in the continuum to unity and shifted vertically for
clarity.} 
\label{fig:f4}
\end{figure*}

In summary, we find  in each of the three  spectra plotted  in Fig. 4 a
broad band with a maximum depth of about 1.5\% of the continuum, and  a central
wavelength close to  7088~\AA\. The differences between these spectra 
give an idea of the  uncertainties involved in the  measurement of this 
band which is traced in the spectra of both telescopes by more than 40
resolution elements at 3$\sigma$  from the continuum, leading to a  very solid
overall detection.   In Fig. 5 we plot overlaid  the three  spectra of the previous figure. The 
combined WHT(a)  and HET (c) spectrum can be fitted using  a
Lorentzian of  FWHM of 40$\pm$5 \AA~ centred at 7088.8 \AA. The 
equivalent width of the interstellar band was measured integrating the
combined spectrum and resulted    W= 600$\pm$200 m\AA~ where the error
takes into account the differences in width and strength of the bands
resulting from the various corrections applied and is largely
dominated by the uncertainty in the location of the continuum. 

\begin{figure*}
\includegraphics[angle=0,width=11cm,height=11cm]{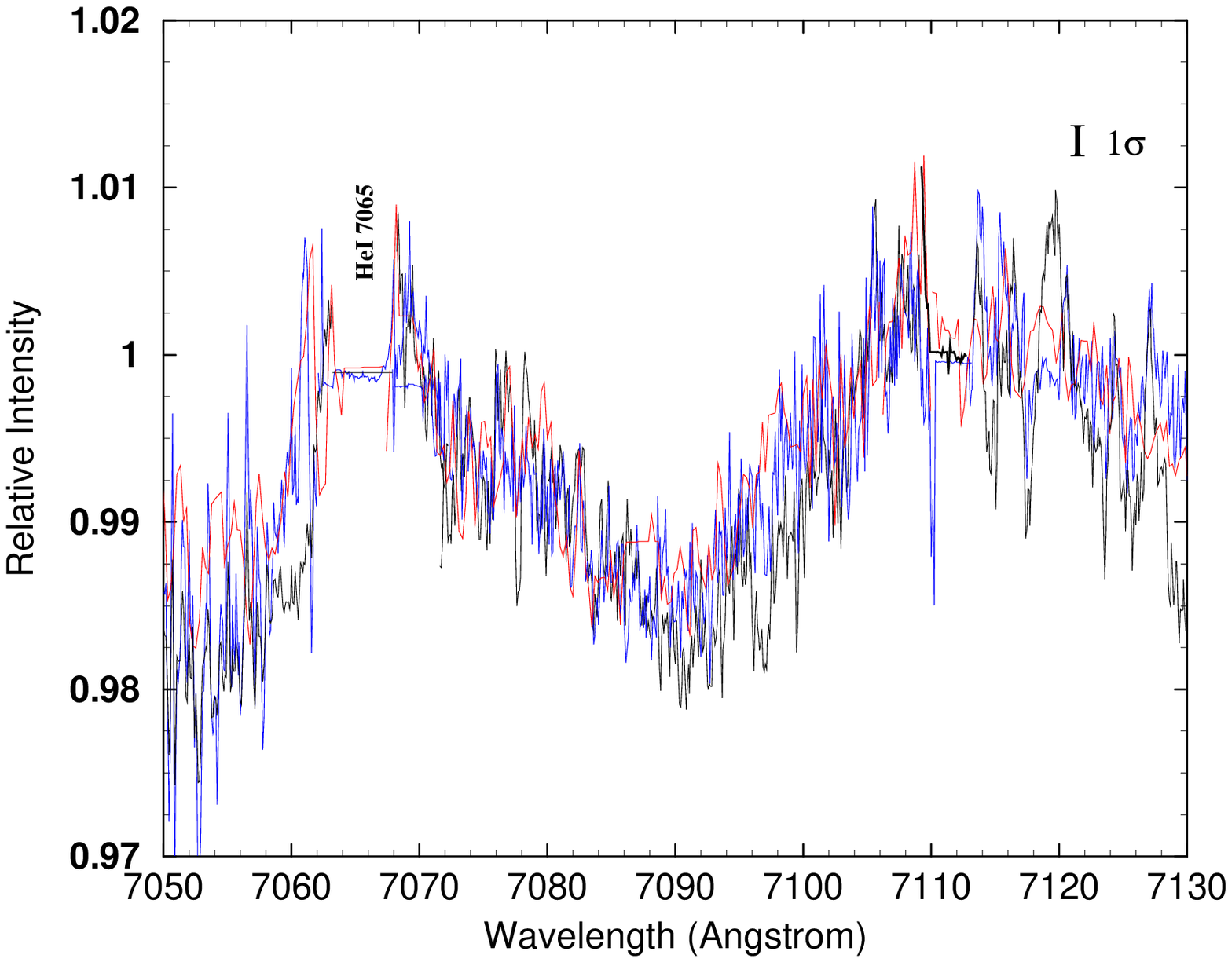}
\caption{Overlay of the HET and WHT spectra in the previous figure.}
\label{fig:f5}
\end{figure*}

The characteristics of the observed 7088 \AA~ band agree reasonably
well with the laboratory measurements (Sukhorukov et al. 2004)  for
the strongest band of the anthracene cation which give  a central
wavelength of 7085.7~$\pm$~1.3~\AA~ and a FWHM of 47 \AA. The FWHM
measured by Sukhorukov et al. in their direct absorption cavity
ring-down spectroscopy seems to be  larger than the observed value of
the interstellar band, but this may well be the result of differences
between  the conditions of their experiment and the conditions of the
cations in the interstellar medium towards Perseus. The study of the
thermal dust emission at millimeter wavelengths indicate temperatures
of order 19 K in the region causing the anomalous microwave emission
(Watson et al. 2005, Tibbs et al. 2010). The laboratory  band of the
anthracene cation was measured when the source temperature was raised
to 205 $^{\circ}$C. According to Sukhorukov et al., the absorption
profile of the anthracene cation  depends on the density and this is
a function of the temperature. In fact at temperatures lower than
120$^{\circ}$C the vapor pressure was too low for these authors to
observe the absorption. They  concluded that in their experimental
setup the anthracene cations may not  experience enough collisions to
cool down the rotational and vibrational degrees of freedom and
therefore that the widths of  laboratory bands are likely  to be 
larger than those of interstellar bands. It is  interesting to note
that with their experimental setup Sukhorukov et al. obtain  larger
widths by a factor 1.5  for the bands of naphthalene cation than
Biennier et al. (2003). The width of the  interstellar band of the
naphthalene cation in this line of sight is claimed by Iglesias-Groth
et al. (2008) to agree  well with the laboratory measurements by
Biennier et al. (2003). This also supports   that the interstellar
band of the anthracene cation is narrower than the current laboratory
band measurements.  

\begin{figure*}
\includegraphics[angle=0,width=11cm,height=11cm]{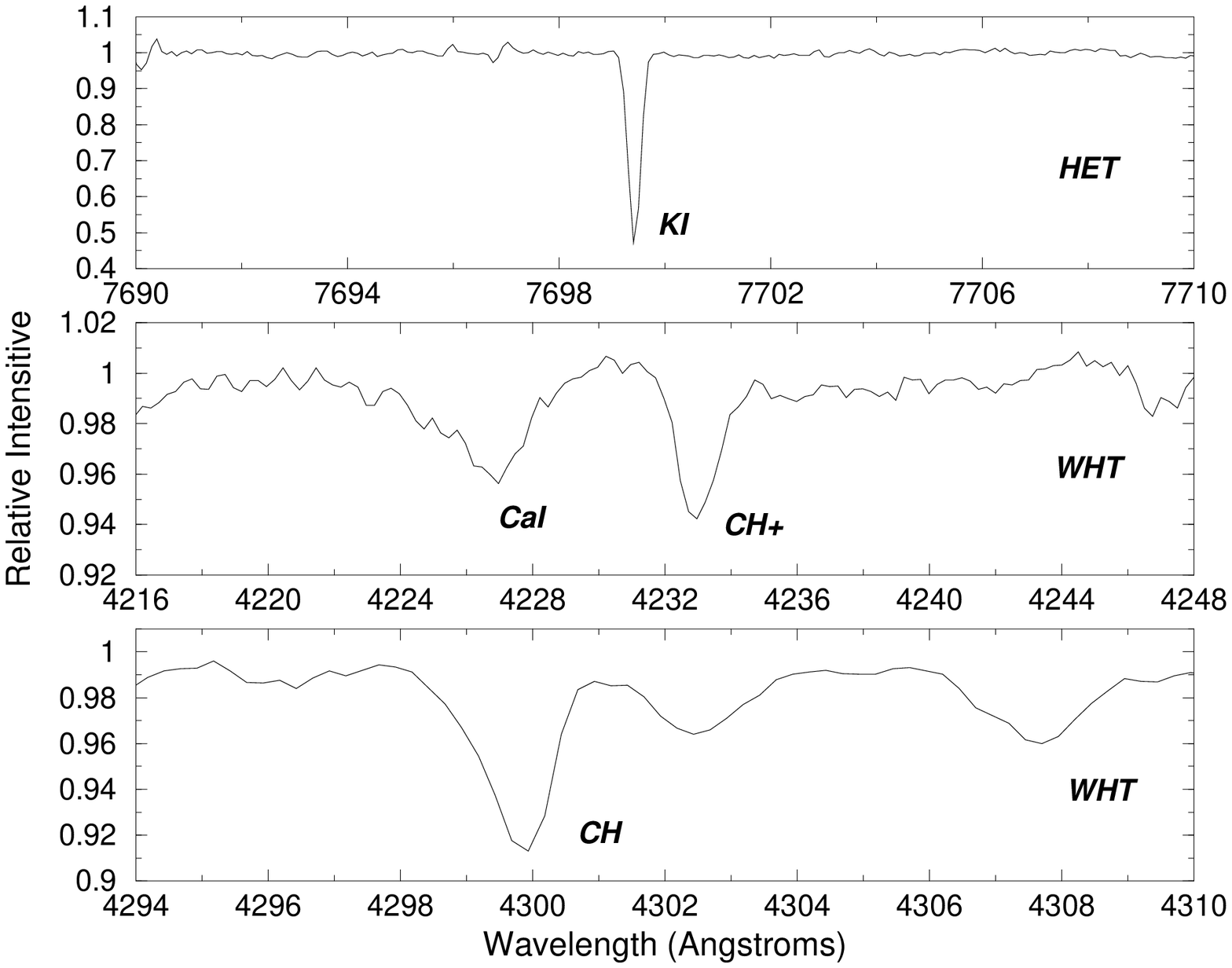}
\caption{Top: HET spectrum of Cernis 52 in the region of the KI line
at 7699 \AA. Middle and bottom: WHT spectra of Cernis 52 in the region
of the  A-X (0-0) bands of  CH$^+$and CH.} 
\label{fig:f6}
\end{figure*}

The 7088~\AA\ band is not only one of the strongest bands from the cation's
ground state but the sole band for which gas-phase spectroscopy has
provided an accurate wavelength.  Matrix isolation spectroscopy (MIS)
(Szczepanski et al. 1993) has provided wavelengths for other bands
but these measurements differ from wavelengths for free cations by amounts
that are too uncertain to provide  a basis for secure identification in an
astronomical spectrum. For example,  the  matrix isolation spectroscopy wavelength for the
7088~\AA\ band is 22~\AA\ to the red of the gas phase measurement.
The 7088~\AA\ band is the 0-0 vibrational band of the 1$^2$2A$_u$ $\leftarrow$ X$^2$B$-{3g}$ transition.
The  0-0 band for 2$^2$A$_u$ $\leftarrow$ X$^2$B$_{3g}$
is at 3520~\AA\ from MIS with an f-value
probably similar to that of 7088~\AA~ (Niederalt et al. 1995).
Then,  3$^2$A$_u$  $\leftarrow$ X$^2$B$_{3g}$ is at 3140~\AA\ from MIS with a predicted f-value
several times
that of the 7088~\AA\ band.  With accurate wavelengths from gas-phase
spectroscopy, it will be of interest to search for these two bands.
Another sequence of transitions involves the sequence of $^2$B$_{1u}$
states. The initial band to 1$^2$B$_{1u}$
is at 6146~\AA\ from MIS but with an f-value
a fraction of the 7088~\AA~ band's f-value it is highly unlikely to be
detectable in our spectrum even if the wavelength were known
accurately.  The next member of the series
to 2$^2$B$_{1u}$ is at 4279~\AA~ from MIS with an
f-value comparable to that for 7088~\AA.
The 0-0 band at 7088~\AA~ (and the 0-0 bands of other electronic transitions)
will be accompanied by a band to the lowest excited vibrational state of the 1$^2$B$_{1u}$
level. For the 7088~\AA\ band, this
band will be about 190~\AA\ to the blue according to the vibrational
frequency suggested by MIS and quantum calculations (Szczepanski et al. 1993). With the f-value suggested by MIS, the band in our
spectrum would have a depth of only about 0.2\%.

In short, attribution of the new interstellar band at 7088~\AA\ 
to the anthracene cation is potentially testable when accurate
laboratory wavelengths are provided for other electronic transitions.
Here we will assume that the 7088.8~\AA\ band is due to anthracene
cation  molecules in the cloud that  causes  the extinction in Cernis
52.  Adopting  for the oscillator strength of the transition at 7088
\AA~  a value of f=0.13 (average of those reported by Hirata et al. 
1993, Niederalt et al. 1995, Ruiterkamp et al. 2005) and using 600
m\AA~  as  measured equivalent width of the band, we derive a column
density of N$_{an^+}$= 1.1 x 10$^{13}$ cm$^{-2}$  (with an uncertainty
of 40\%). This is very  similar to the column density estimated for
the naphthalene cation in the same line of sight (Iglesias-Groth et
al. 2008).  Assuming a ratio C/H= 3.7 x 10$^{-4}$ and hydrogen column
density per unit interstellar reddening from Bohlin et al. (1978) we
obtain N(H)= 5.3 x 10$^{21}$ cm$^{-2}$  and derive that  $\sim$ 0.008
\%  of the total carbon in the intervening cloud could be in the form
of anthracene cations.    

With these low abundances, electric dipole emission of  these molecules can explain only a small fraction of the anomalous microwave emission detected in the line of  sight, mainly at  frequencies higher
than 50 GHz (Iglesias-Groth et al. 2008). It is possible that neutral PAHs are more abundant than cations and that larger PAHs  exist in the cloud and contribute  to the anomalous emission at frequencies in the range 10-50 GHz. 

\begin{figure*}
\includegraphics[angle=0,width=11cm,height=11cm]{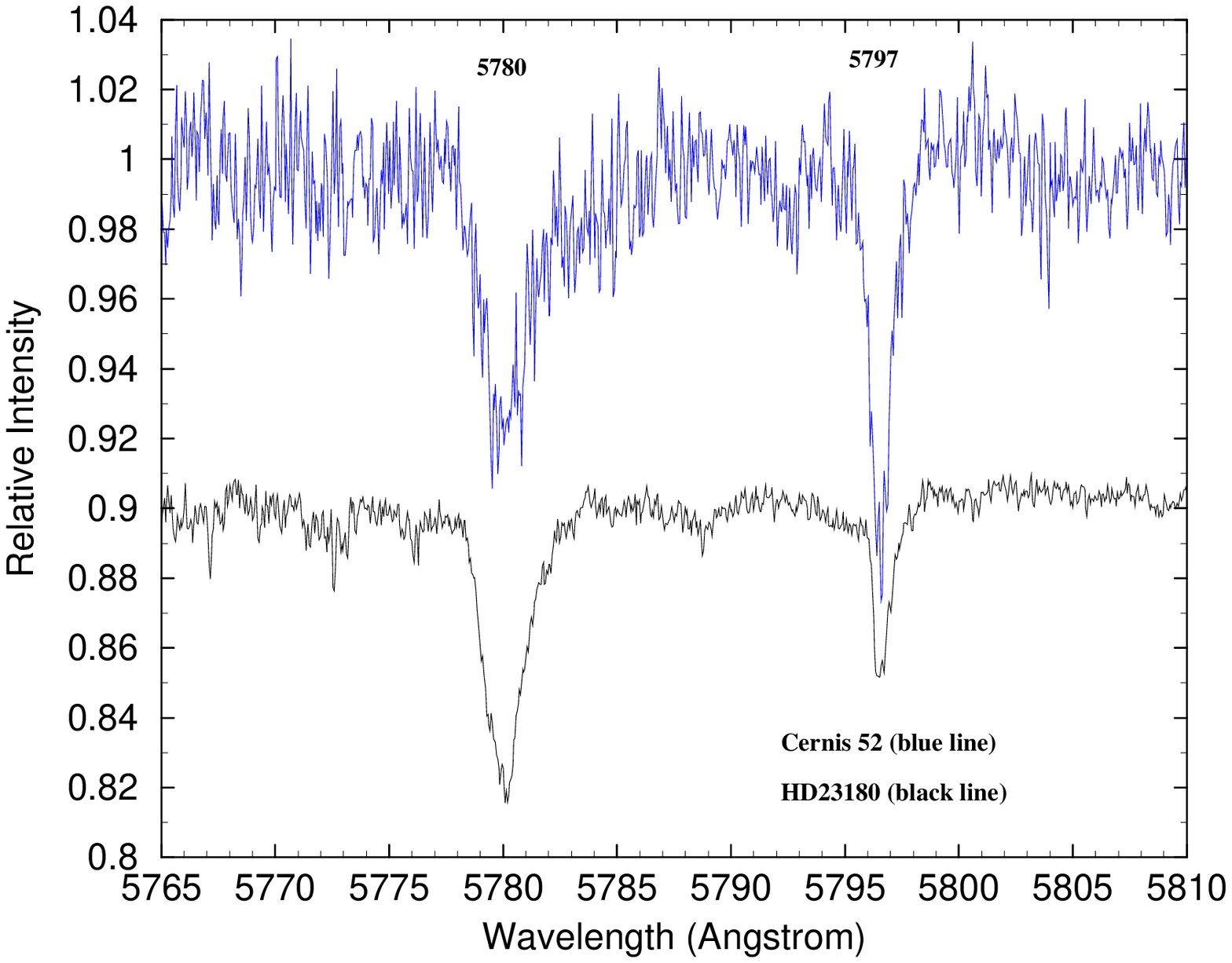}
\caption{Spectra of Cernis 52 (blue line) and HD 23180 (black line) in
the region of the DIBs at 5780 and 5797 \AA.}. 
\label{fig:f7}
\end{figure*}

\subsection{The intervening cloud}

We note the existence of a reflection nebula in the line of sight of
Cernis 52, clearly seen as extended optical emission in images of the
Digital Sky Survey.  The radial velocity of the star   and of the DIBs
in the spectrum agree within 1 km/s, the uncertainty in the
measurement, thus this  star which is a likely  member of the IC 348
cluster (Gonz\'alez Hern\'andez et al. 2009)  could  be embedded in
the cloud. Observations of the K I 7698.974 \AA~ line, show only one 
velocity component, indicative of possibly a single cloud structure in
this line of sight (see Fig. 6).   Extended mid-infrared emission is
also seen  in archive images obtained  with  IRAC onboard the Spitzer
satellite in the wavelength range 3 to 9  $\mu$m.  This emission  is
much stronger at the 5.8 and 8.0 $\mu$m bands than at 3.6 and 4.5
$\mu$m. We attribute the high emission at  long wavelengths to C-C and
C-H stretching and bending modes of  PAHs  at 6.2 and 7.7+8.6 $\mu$m,
and note that the weaker emission of the  expected 3.3 $\mu$m band
could be due  to a significant ionization of PAHs (Iglesias-Groth et
al. in preparation).    

In order to provide some insight on the physical and chemical conditions and in particular on the
PAH charge state distribution	of the cloud in the line of sight of Cernis 52, we have measured
the interstellar lines of CH and CH$^+$ at 4300.313 \AA~ and 4232.548 \AA~ available in our ISIS
blue arm spectra taken in December 2007 and 2008 at WHT.  These lines have been  used in the
literature to  study  a large variety of interstellar (diffuse, translucent, dense) clouds 
including several with  extinction similar to that of the cloud under consideration here (see e.g.
de Vries and van Dishoeck 1988; Crane, Lambert and Sheffer 1995).  We find the blue wing profile
of the CH line slightly asymmetric possibly due to contamination by a stellar photospheric
feature. We used  a synthetic stellar spectrum to take into account  potential contributions of
photospheric lines from Cernis 52 and correct for them. The CH$^+$ line appears consistent with a
single Gaussian profile.  We measure  equivalent widths of   W=66$\pm$8 m\AA~ and W=70$\pm$8
m\AA~ for the CH and CH$^+$ lines, respectively.  These  errors take into account the
uncertainty in the measurement of the equivalent widths which is  of order 5-6 m\AA~ and  the
uncertainty associated with the correction of photospheric lines which is driven by  the
uncertainties in the stellar parameters of Cernis 52 (Gonz\'alez Hern\'andez et al. 2009). We
estimate uncertainties of  order 20\% for  the predicted  equivalent widths of the photospheric
lines  which in both cases are minor contributors to the absorption at the wavelength of the CH
features.  This  imply additional  uncertainties of order 4-5 m\AA~
which combined quadratically with the previous error lead to the final
quoted 8 m\AA~ error. 
The two sets of WHT  observations taken with a time baseline of one
year gave  consistent  strengths and  the observed wavelengths of each
set of  lines  agreed  within 50 m\AA~ (4 km s$^{-1}$). Both  CH and
CH$^+$ lines  show  heliocentric velocities consistent within the
uncertainties  with that of the  KI line observed at higher dispersion
with  HET. The oscillator strength of the CH transition is  f=0.00506
(Larsson and Siegbahn 1983, Brzozowski et al. 1976). In order to
account for a slightly saturated line, the  column density will be
obtained from curves of growth with b=1.5 km s$^{-1}$. The measured
equivalent width implies then	N(CH)=(2$\pm$0.2) x 10$^{14}$
cm$^{-2}$. Similarly, and  adopting the  oscillator strength of the
CH$^+$ line from Larsson and Siegbahn (1983b) we obtain N(CH$^+$)=
(2$\pm$0.2) x 10$^{14}$ cm$^{-2}$. Both species display rather high
column densities.

The average HI column density derived from the LAB map (Karberla et
al. 2005) in the direction of Cernis 52  is N(HI)= 1.2 x 10 $^{21}$
cm$^{-2}$. Using  the relationships between CH and H$_2$ in Danks,
Federman and Lambert (1984) and Mattila (1986)   we  obtain from our  
N(CH) value two independent estimates of the  column density for
molecular hydrogen. The  average value results N(H$_2$)=6.3 x
10$^{21}$ cm$^{-2}$. The total  column density of hydrogen  then is 
N(H)=N(HI)+ 2N(H$_2$)=13.8 x 10 $^{21}$ cm$^{-2}$. This is a factor
2.6 larger  than the value  derived from the excess color  E(B-V)   of
Cernis 52.  Possibly indicating that  the A$_V$ and E(B-V) relation is
steepened in this cloud. It seems that  we deal with a cloud  that
causes a total visual  extinction  significantly higher than in
typical diffuse interstellar clouds and  comparable or even higher
than   translucent molecular  clouds (see e.g. van Dishoeck and Black
1989). Models for clouds of similar  characteristics  can be found for
example in Ruiterkamp et al. (2005) who include  a detailed treatment
of  the PAH charge state distribution. It is apparent  from their Fig.
3 that PAHs with a low number of carbon atoms are mostly expected in a
neutral stage, in particular  anthracene and naphthalene cations 
charge fractions are expected  of order 10\% while the neutral species
could be as abundant as 65\%  and  anions would provide the remaining
fraction. Changes in the hydrogen density (in the recombination rate)
or in the intensity of the UV radiation field (in the ionization rate)
do not affect drastically this result for this type of cloud. 

The derived  column densities of CH and CH$^+$ in the cloud toward
Cernis 52 are significantly higher than in any other CH surveys of
diffuse interstellar 
clouds, including examples of clouds of higher extinction. The high
abundance of these species and also of molecular hydrogen is an
indication of a potentially rich 
chemistry that may convert  atoms into organic molecules (see e.g.
Millar 2004 for the reaction networks involved) at the observed column
densities toward Cernis 52. In less rich clouds previously surveyed in
the literature the column density of anthracene and naphthalene  may
be  considerably lower and their bands weaker. This could explain why
the anthracene and naphthalene cation bands discussed here and in
Iglesias-Groth et al. (2008)   have escaped detection in other lines
of sight. Another reason for the lack of previous detections of the
anthracene cation is that DIB surveys are frequently conducted using  
high dispersion echelle spectrographs which have difficulties to
detect weak very broad features due to continuum correction. 

We have  examined the behaviour of the major DIBs in the various spectra  of Cernis 52 available to us (including those obtained by  Iglesias-Groth et al. 2008 at the 3.6m Telescopio Nazionale Galileo (TNG) and
the 2.7m at Mc Donald Observatory) with the aim to identify any possible anomalous behaviour which could be related to the dust properties of the cloud. Here, we  present results for the two major diffuse bands at
5780 and 5797 \AA.  These two diffuse  bands discovered by Heger (1922) show strength ratios that vary with the shape of the extinction curve (Krelowski et al. 1987). High molecular abundances - as  for CH in the
line of sight of Cernis 52 -  appear in interstellar clouds  characterized by a broad UV extinction bump (2175 \AA), steep far-UV extinction and low 5780/5797 diffuse band intensity ratio. Such clouds are called
``zeta'' type clouds (Krelowski and Sneden, 1995) and present strong molecular features in their spectra. Examples of these clouds are found towards stars like $\zeta$ Per and HD 23180 in Perseus where the
5780/5797 ratios (equivalent widths of 5780 to 5797) range between 1 and 2 (Krelowski et al. 1996). Such  values are much lower than those found by these authors in the so-called ``sigma'' clouds which present
strikingly  different extinction curves (see for a discussion  Krelowski, Sneden and Hiltgen, 1995).    In Fig. 7, we plot our Mc Donald spectra of Cernis 52 in the spectral range of these DIBs in comparison with
HD 23180.  It is remarkable the similar strength of the 5780 DIB in both stars while the 5797 DIB is clearly stronger in Cernis 52. This leads to a smaller 5780/5797 ratio  than in HD 23180  and  fully  supports
the classification of the intervening cloud towards Cernis 52  as of ``zeta'' type. To our knowledge the UV  extinction curve in this line of sight is unknown but based in the behaviour of the 5780/5797 DIBs  we
may expect a steep far-UV extinction. It is well established  that some diffuse bands like the 5797 \AA~ correlate positively with the overall slope of the extinction curve while others like the 5780 \AA~ show
negative correlation (Megier, Krelowski and Weselak 2005). A postive correlation with the overall slope probably indicates an anticorrelation with the UV irradiation and the carrier of the 5797 \AA~ band may
benefit from shielding. If this is the case shielding is more effective in the Cernis 52 cloud than in the ``zeta''  cloud towards the  star HD 23180 (also in Perseus but in a region without significant anomalous
microwave emission). The band at 5780 \AA, known to be more resistant to strong UV fields than the 5797 DIB, shows similar strength in both clouds.  The DIB at 5850 \AA~  is  known to follow closely the behaviour
of the 5797 DIB and correlates with the the overall slope of extinction (Megier et al. 2005). We find that the strength of this DIB  in Cernis 52 is also  a factor two higher than in HD 23180. The extinction in
the UV reduces the flux of the UV radiation that is believed to ionize or destroy the carriers of these bands.   

We have scrutinized our  spectra to search for bands of other PAHs with
reliable gas phase  measurements (see Salama 2008) and confirm  in our new data
the naphthalene band at 6707 \AA~previously found by Iglesias-Groth et
al. (2008). At various wavelengths  we find marginal evidence for
other  broad bands that may correspond to PAH cations, and for
relatively narrow absorptions that could be associated to  carbon
chains and carbon rings   but  confirmation  may  require higher
quality measurements. The upper limits we can impose on  the column
densities of these other PAHs do not bring at this stage sufficient
information to establish the reaction networks in the cloud. A
comprehensive analysis of the DIBs in the line of sight of Cernis 52 
may also  provide additional information on the UV radiation field in
the intervenig cloud  but this is  beyond the scope of this paper and 
will be subject of a dedicated more comprehensive  work. 

\section{Conclusions}

We report the discovery of a new  broad band at 7088.8$\pm$2 \AA~ in
the spectrum of Cernis 52  with wavelength and width consistent with
gas phase measurements of the strongest band of the anthracene cation.
This is the  only band of the anthracene cation  with a wavelength
measured from gas phase spectroscopy. We measure an equivalent width
of W=600$\pm$ 200 m\AA. Assigning the observed band to this cation, we
infer a column density of N$_{an^+}$= 1.1($\pm$0.4) x 10$^{13}$
cm$^{-2}$. This value is very close    to previous estimates for the
column density of naphthalene cations in the same line of sight
N$_{naph^+}$=1.0($\pm$0.6) x 10$^{13}$ cm$^{-2}$ (Iglesias-Groth et
al. 2008). It appears that the intervening cloud contains the same
amount of cations of anthracene and naphthalene.  We also detect bands
of CH and CH$^+$ at 4300.313 and 4232.548 \AA~  from which we obtain
column densities for these species of 2($\pm$0.2)x 10$^{14}$ cm$^{-2}$
and  infer a rather high column density for molecular hydrogen
N(H$_2$)=6.3 x 10$^{21}$ cm$^{-2}$.  Observations  of the KI line are
consistent with  the existence of a single cloud in the line of sight.
Measurements of the major DIBs at 5780 and 5797 \AA~ clearly indicate
that the intervening cloud is of ``zeta'' type, probably with a steep
far-UV extinction.  

The cloud toward Cernis 52 is likely to be responsible for the
anomalous microwave emission reported in this line of sight. The
presence of PAHs in a region of strong anomalous microwave emission
adds support to the hypothesis that electric dipole radiation by these
type of  molecules  is responsible for the  excess emission detected
in the 10-50 GHz range. The  abundance of PAHs was taken as  a
free-parameter in  models attempting to explain this anomalous
microwave emission (Draine and Lazarian 1998). The abundances derived
for naphthalene and anthracene cations provide additional constraints
for  these models.    

Laboratory experiments at low temperatures (15 K) have shown that
UV radiation on a mixture of   naphthalene,  ammonia and  H$_2$O
ice leads to the formation of a large variety of amino acids (Chen
et al. 2008). In the line of sight of  Cernis 52 there is evidence
for amonnia  (Rosolowsky et al. 2008) with a column density quite
similar to what we find for the most simple PAHs. The  likely
presence of H$_2$O ices combined with  the UV radiation from Cernis
52  could make this region in  Perseus  a  potential factory for
amino acids. The detection of specific PAHs  in the molecular cloud
towards Cernis 52 may help to identify  a possible path for
biogenic compounds from the chemistry of clouds (where new stars
form) to the pre-biotic molecules detected  in protoplanetary
disks. Further studies in this remarkable region  have the
potential to reveal a very rich interstellar prebiotic chemistry. 


\section*{ACKNOWLEDGEMENTS} 
We acknowledge support front grant AYA-2007-64748 from the Spanish
Ministry of Science and Innovation and also we thank the Mc Donald
Observatory's Hobby-Eberly Telescope time allocation committee for
their exceptional support. DLL thanks the Robert A. Welch Foundation
of Houston for support via grant F-634. J.I.G.H. thanks financial
support from the Spanish Ministry project MICINN AYA2008-00695.
D.A.G.H. acknowledges support from the Spanish Ministry of Science and
Innovation (MICINN) under the 2008 Juan de la Cierva Programme.


\begin{thebibliography}{}

\smallskip
\bibitem[]{Allama89} Allamandola, L. J., Tielens,G. G. M. \& Barker, J.R., 1989, \apjs, 71, 733,1989
\smallskip
\bibitem[]{Allende04} Allende Prieto, C., Barklem, P.S., Lambert, D.L., Cunha, K. 2004, A\&A 420, 183
\smallskip
\bibitem[]{Biennier03} Biennier, L., et al. 2003, J. Chem. Phys., 118, 7863
\smallskip
\bibitem[]{bohlin} Bohlin, R.C., Savage, B.D. \& Drake, J.F. 1978, \apj, 224, 132 
\smallskip
\bibitem[]{brzozowski76} Brzozowski, J., Bunker, P., Elander, N~\& Erman, P.,1976, \apj, 207, 414
\smallskip
\bibitem[]{Casassus06} Casassus, S.; Cabrera, G. F., F\"{o}rster, F., Pearson, T. J., Readhead, A. C. S.\& Dickinson, C., 2006, \apj~639, 951
\smallskip
\bibitem[]{cernis93} Cernis, K. 1993, Baltic Astronomy 2, 214
\smallskip
\bibitem[]{chen08} Chen, Y.-J., Nuevo, M., Yih, T.-S., Ip, W.-H., Fung, H.-S., Cheng, C.-Y., Tsai, H.-R.~\& Wu, C.-Y. R., 2008, MNRAS, 84, 605
\smallskip
\bibitem[]{crane95}Crane, P., Lambert, D.L. \& Sheffer,Y., 1995, \apjs, 99,107
\smallskip
\bibitem[]{danke84} Danks, A.C., Federman, S.R. \& Lambert, D.L., 1984, A\&A,130, 62
\smallskip
\bibitem[]{drain98a} Draine, B.T. \& Lazarian, A. 1998, \apjl, 494, L19
\smallskip
\bibitem[]{deOliveira99,02} de Oliveira-Costa, A., et al. 1999, \apjl~, 527, L9
\smallskip
\bibitem[]{deOliveira99,02} de Oliveira-Costa, A., et al. 2002, \apj, 567, 363
\smallskip
\bibitem[]{deVries88} de Vries, C.P. \& van Dishoeck, E.F., 1988, A\&A, 203, L23
\smallskip
\bibitem[]{galaz01} Galazutdinov, G.A., Musaev, F.A., Kre{\l}owski, J. \& Walker, G.A.H. 2000, PASP, 112, 648
\smallskip
\bibitem[]{jonay09} Gonz\'alez Hern\'andez,J.I., Iglesias-Groth, S., Rebolo, R., Garc\'\i a-Hern\'andez, D.A., Manchado, A. \& Lambert, D.L., 2009, \apj~706, 866 
\smallskip
\bibitem[]{heger22} Heger, M.L. 1922, Lick Obs. Bull., 10, 146
\smallskip
\bibitem[]{Hilde07} Hildebrandt,S. R., Rebolo, R., Rubiño-Mart\'\i n, J.A., Watson, R.A., Guti\'errez, C.M., Hoyland, R.J. \& Battistelli, E.S. 2007, MNRAS 382, 594
\smallskip
\bibitem[]{Hirata99} Hirata, S., Lee, T.L. \& Head-Gordon, M. 1999,  J. Chem. Phys. 111, 8904
\smallskip
\bibitem[]{hobbs08} Hobbs, L. M., York, D. G., Snow, T. P., Oka, T., Thorburn, J.A.; Bishof, M., Friedman, S. D. \& McCall, B. J. et al.2008, \apj~680, 1256
\smallskip
\bibitem[]{ig0506} Iglesias-Groth, S. 2005, \apjl~ 632, L25; 2006, MNRAS 368,1925
\smallskip
\bibitem[]{ig08} Iglesias-Groth,S., Manchado, A., Garc\'\i a-Hern\'andez, D.A., Gonz\'alez Hern\'andez, J.I. \& Lambert, D.,2008, \apjl~ 685, L55 
\smallskip
\bibitem[]{karberla05} Kalberla, P.M.W., Burton, W.B., Hartmann, Dap, Arnal,E.M., Bajaja, E., Morras, R., and P\"{o}ppel, W.G.L., 2005, A\&A, 440, 775
\smallskip
\bibitem[]{krelowski87}Krelowski, J., Walker G.A.H., Grieve, G.R., Hill, G.M. 1987, ApJ, 316, 449
\smallskip
\bibitem[]{Krelowski95}Krelowski, J. \& Sneden C. 1995, in :''The Diffuse Interstellar Bands'' IAU Coll. 137, A.G.G.M. Tielens and T.P. Snow (eds.) Kluwer, Dordrecht, p. 13
\smallskip
\bibitem[]{Krelowski96}Krelowski, J., Megier, A. \& Strobel, A. 1996,  A\&A, 308, 908
\smallskip
\bibitem[]{kogut96} Kogut, A., Banday, A. J., Bennett, C. L., Gorski, K. M., Hinshaw, G.\& Reach, W. T. 1996 ApJ, 460, 1
\smallskip
\bibitem[]{larsson83}Larsson, Mats \& Siegbahn, Per E.M., 1983, Chem. Phys 79, 2270
\smallskip
\bibitem[]{larsson83b}Larsson, M.; Siegbahn, P. E. M., 1983b, Chem Phys 76, 175
\smallskip
\bibitem[]{leitch97} Leitch, E. M., et al., 1997, \apjl~ 486, L23
\smallskip
\bibitem[]{Niederalt95} Niederalt, C., Grimme, S., Peyerimhoff, S. D., 1995,Chem.Phys.Lett., 245, 455
\smallskip
\bibitem[]{mattila86} Mattila, K, 1986, A\&A, 160, 157
\smallskip
\bibitem[]{millar04chap1}Millar, T.J, 2004,Organic Molecules in the InterstellarMedium  in Astrobiology: Future Perspectives, eds. P. Ehrenfreund et al.,Kluwer, 17-31.
\smallskip
\bibitem[]{puget89} Puget, J. L. $\&$ L\'eger, A. 1989, ARA~\& A  27,161
\smallskip
\bibitem[]{Rosolowsky08} Rosolowsky, E. W., Pineda, J. E., Foster, J. B., Borkin, M. A., Kauffmann, J.,Caselli, P., Myers, P. C.\& Goodman, A. A.,2008 \apjs~ 175, 509
\smallskip
\bibitem[]{Ruiterkamp05} Ruiterkamp, R., Cox, N. L. J., Spaans, M., Kaper, L., Foing, B. H., Salama, F.\&~Ehrenfreund, P., 2005, A\&A, 432, 515
\smallskip
\bibitem[]{salama08iau} Salama, F., 2008,  IAU Symp. 251, 357.          
\smallskip
\bibitem[]{Sukhorukov08} Sukhorukov,O., Staicu, A., Diegel, E., Rouill\'e, G., Henning, Th. \& Huisken, F.,2004, Chem. Phys. Lett. 386, 259
\smallskip
\bibitem[]{tibbs09} Tibbs C. et al. 2010,  MNRAS 402, 1969
\smallskip
\bibitem[]{vanDishoeck89} van Dishoeck, Ewine F.$\&$ Black, John H., 1989, \apj~340, 273
\smallskip
\bibitem[]{watson05,06} Watson,R.A. et. al 2005,  \apj~ 624, L89
\end{thebibliography}
\bibliographystyle{mn2e}

\end{document}